\documentclass{article}

\usepackage{arxiv}

\usepackage[utf8]{inputenc} 
\usepackage[T1]{fontenc}    
\usepackage{hyperref}       
\usepackage{url}            
\usepackage{booktabs}       
\usepackage{amsfonts}       
\usepackage{nicefrac}       
\usepackage{microtype}      
\usepackage{lipsum}
\usepackage{graphicx}
\usepackage{float}
\usepackage{xcolor}
\usepackage{hyperref}

\title{Characterization of Political Polarized Users Attacked by Language Toxicity on Twitter}

\author{
 Wentao Xu \\
  Department of Science and Technology of Communication\\
  University of Science and Technology of China \\
  \texttt{myrainbowandsky@gmail.com} \\
  {\color{red} \textbf{This work is accepted by CSCW 2024: \url{https://doi.org/10.1145/3678884.3681849}}}
  }
 
\begin{document}
\maketitle
\begin{abstract}
Understanding the dynamics of language toxicity on social media is important for us to investigate the propagation of misinformation and the development of echo chambers for political scenarios such as U.S. presidential elections.
Recent research has used large-scale data to investigate the dynamics across social media platforms. 
However, research on the toxicity dynamics is not enough. 
This study aims to provide a first exploration of the potential language toxicity flow among Left, Right and Center users.
Specifically, we aim to examine whether Left users were easier to be attacked by language toxicity.
In this study, more than 500M Twitter posts were examined. It was discovered that Left users received much more toxic replies than Right and Center users. 
\end{abstract}


\section{Introduction}

Social media has become an indispensable daily element of contemporary social life \cite{doi:10.1177/1745691612442904}.  
When social media platforms bring people freedom of communication, studies identified language toxicity emerging across platforms, such as Twitter (Now,  X)~\cite{10.1145/3041021.3053890,10.1145/2818052.2869107}, Facebook~\cite{doi:10.1177/1527476420982230,10.1145/3232676}, Reddit~\cite{10.1145/3543507.3583522}, YouTube~\cite{Miyazaki2024}, Telegram~\cite{Wich_Gorniak_Eder_Bartmann_Çakici_Groh_2022}, and Whisper~\cite{Silva_Mondal_Correa_Benevenuto_Weber_2021}.
These studies have identified various forms of toxic content, including violence, obscenity, threats, insults, and abusive language. 
Toxic language in online social networks is prevalent between users with no connection than between mutual friends, and mildly offensive terms are used more frequently to express hostility between these two groups\cite{Radfar_Shivaram_Culotta_2020}.
The nature and extent of toxic language can vary by platform. For instance, Reddit has been found to contain a higher frequency of posts with insults, identity attacks, threats of violence, and sexual harassment\cite{Kumar2023}. 
Additionally, the research indicates that while most studies on offensive language detection have focused on English, the toxic contents have been identified in other languages, such as Greek and Indonesian, as well \cite{pitenis-etal-2020-offensive, OkkyIbrohim_2019}. 
In addition, the presence of toxic language on Twitter can have significant negative impacts on individuals and communities, even affecting mental health\cite{Rost2016}.

In political scenarios, online conversations during U.S. presidential elections can indeed exhibit toxicity~\cite{Ventura_Munger_McCabe_Chang_2021}. 
While social media platforms are praised for enhancing democratic discussions, the presence of social bots can distort political discourse, potentially influencing public opinion and election integrity negatively\cite{doi:10.1177/20563051211008827,Bessi2016}. The behaviour of social bots aggravates the propagation of toxic content. The toxicity in online political talk is often linked to incivility, challenging the perception that it is beneficial for the elections.
Moreover, the study of online chatter surrounding elections is crucial for ensuring evidence-based political discourse and free and fair elections~\cite{Chen2021}.
Therefore, while online conversations can provide a platform for political discussions, the toxicity and manipulation through bots underscores the importance of monitoring and studying these interactions to safeguard the election process.

However, the political process is severely affected by polarization.
Polarization is popular on Twitter, as the platform serves as a significant space for political discourse, which can influence public opinion and democratic processes.
Studies have shown that Twitter can both facilitate cross-ideological exchanges and contribute to the clustering of users around shared political views, potentially reinforcing partisan loyalties and contributing to polarization~\cite{https://doi.org/10.1002/1944-2866.POI354}.

The impact of Twitter on political polarization is also significant in fragmented political systems, where the platform's role in shaping communication among political entities can affect collaboration between parties and the overall political landscape~\cite{Borah2022}. Interestingly, while some research supports the ``echo chambers'' view, suggesting that social media platforms like Twitter foster political polarization by creating fragmented and niche-oriented spaces for like-minded individuals, other studies highlight the presence of cross-cutting interactions that could mitigate this effect~\cite{HONG2016777}. 

Moreover, the influence of social media, on political polarization has been demonstrated through simulations, indicating that the type of political views presented in social media can shape the political orientation of a population~\cite{Goyal_Goyal_2023}. The investigation into political polarization on Twitter is crucial due to the platform's ability to shape political communication and influence public opinion. While there is evidence of both echo chambers and cross-ideological interactions, the overall effect of Twitter on political polarization varies and is a subject of ongoing scholarly debate. 

Language on Twitter can potentially exacerbate political polarization \cite{10.1007/978-3-030-82147-0_5}. 
The language of toxicity can be bared in the echo chambers on Twitter, leading to more extreme and toxic language, reinforcing divisions\cite{HONG2016777}. Moreover, language on Twitter can also be affected by traditional media, such as broadcast media language, which contributes to toxic online interactions\cite{Ding_Horning_Rho_2023}. 
Additionally, Twitter is vulnerable to manipulation by malicious actors who use polarized and toxic language to sow discord. This can significantly distort the political landscape and influence public opinion~\cite{10.1093/pnasnexus/pgac019}.

The most recent extensive research on language toxicity shows that toxicity does not always increase as online discussions progress, suggesting more rounds of conversation may not lead to higher toxicity~\cite{Avalle2024}.
However, the detailed dynamics of these toxic politically polarized conversations are still unclear. Here we focus on the replies to politically polarized users on Twitter.
In most cases, any social networking service (SNS) account can freely engage with another one with texts, which could lead to further negative and harmful online social behaviours, such as rounds of toxic replying.
\cite{Miyazaki2022} discovered that anti-vaxxers are more aggressive in replying by analyzing toxic replies of English and Japanese tweets.
\cite{mosleh2022measuring} found that ideological extremity is more associated with the conservatives than the liberals through network analysis.
\cite{10.1145/3442381.3449861} identified toxic replies diffusing patterns on Twitter based on news outlets diffused on Twitter.
These studies helped understand the mechanism of replies on Twitter, but the language toxicity patterns of politically polarized Twitter replies were not investigated.
This study examined the correlation between political polarization and language toxicity of Right, ``Center,'' and Left replies.

\section{Data \& Methods}

\subsection{Data}
It is well known that the COVID-19 pandemic is a worldwide healthcare crisis, during which political polarization was intensified \cite{doi:10.1073/pnas.2216179120,10.3389/fpos.2021.622512,doi:10.1073/pnas.2117543119}.
Such a catastrophic global situation provides a time window to examine the association between political polarization and online language toxicity.

In this study, 542,212,429 English tweets were collected from February 20 2020 to May 30 2022 by querying COVID-19-related keywords: ``corona virus'', ``coronavirus", ``covid19'', ``2019-nCoV'', ``SARS-CoV-2'', ``wuhanpneumonia'' using the Twitter Search API.
A total of 25,370,268 replies of English tweets were used for this study.

\subsection{User annotation}
A politically-leaning URL domain list of news websites was then obtained by requesting from Allsides\footnote{www.allsides.com} for academic research purposes, which contains 160 Left and Lean Left URLs, 98 Right and Lean Right URLs and 180 Center URLs.
Based on the list, each reply was labelled as Right if its domain of the Twitter URL object was identified in the Right or Lean Right domain list; the other replies were labelled as Left and Center, accordingly.

To examine the degree to which a user engages with labelled replies, we categorized users according to their replies' domain labels. 
For example, the Right user category includes users whose reply URL objects contain the Right domains, exclusively. 
It happens that a reply does not contain any URLs.
Please, keep in mind that this study only looked at replies that met two criteria:
\begin{itemize}
\item {The user to whom were replied (``in\_reply\_to\_screen\_name'' in the standard Twitter object \footnote{\url{https://developer.twitter.com/en/docs/twitter-api/v1/data-dictionary/object-model/tweet}}) is not a ``Null'' value.}

\item {The reply contains at least, one domain in the Twitter URL object.}

\end{itemize}
Meanwhile, the study further considered the frequency with which each user was replied to in each politically-leaning category. 
For example, if the Left domains occurred in a replied-to user's reply URL object three times without Center and Right domains occurring, this user was considered to be a three-time-replied-to user in the Left category and was called a \emph{three-time-replied-to Left user}. 
As a result, a user who was replied more frequently, in this study, is regarded as a more engaged user with a politically-leaning domain category.
To this end, Twitter users were annotated as Left, Right and Center categories.

\subsection{Toxicity Calculation}
The Perspective API~\footnote{\url{https://www.perspectiveapi.com/}} is considered suitable for toxicity calculation due to its machine learning-based approach to detecting and moderating toxic content on social media platforms~\cite{doi:10.1177/20539517211046181,pozzobon-etal-2023-challenges,Sharma2023, Schramowski_2022,Miyazaki2022,Miyazaki2024,mosleh2022measuring}. It has been adopted for content moderation, monitoring, and research purposes.
It aligns well with human ratings of toxicity~\cite{Avalle2024} and disrespectfulness, especially for highly toxic comments~\cite{rosenblatt-etal-2022-critical}, indicating the capability of language toxicity measurement for Perspective API is robust.

For the text input into the Perspective API, a probability score scaling within $[0,1]$ is calculated. 
The higher the score is, the more toxic the input text is.
Some research uses a threshold for classifying ``toxic'' and ``nontoxic'' texts.
Here, this strategy was not adopted, as I need to characterise the toxicity of all users. 
To measure the toxicity of each user, the replied texts for each user were aggregated, and then sent to Perspective API. 
Since each category of users possesses various statistical indicators for toxicity, here, the analysis for maximum and median toxicity scores of Left, Right and Center was reported.

\section{Results}

\subsection{The Left received much more toxic replies.}

\begin{figure}[H]
  \centering
  \includegraphics[width=0.5\linewidth]{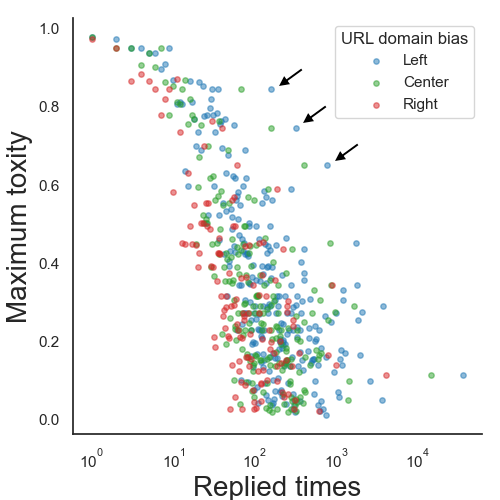}
  \caption{The maximum toxicities of replies of each politically-leaning category of replied-to users received.
  The X-axis,repied times, is in log scale. The Y axis indicates the times each category of users were replied. Reds indicate users were engaging with Right domains, exclusively; Greens indicate users were engaging with Center domains, exclusively; Blues indicate uses were engaging with Left domains, exclusively. Top toxic to-replied users were indicated by arrows. Left category outliers were indicated by arrows. Replies to the Left category users were significantly more toxic than the ones to the Right and Center category ($p < 0.005$ by Mann–Whitney U-test with a Bonferroni correction.).}
  \label{fig1}
\end{figure}

The overall negative correlation between maximum toxicity
and the replied times was identified in this study(Figure.\ref{fig1}).

The maximum toxicity is the highest value of language toxicity of the category with specific replied times.
Figure 1 illustrates the maximum toxicities of each category replied at different times, indicating that more-replied-to users were less likely to receive replies with
high toxicity.
The Right and Center categories replied-to users shared a similar maximum toxicity distribution (Kolmogorov-Smirnov test, $p>0.05$), while the Left category showed a different distribution ($p<0.05$).
However, the statistical difference does not change the overall trend of the three categories.

\begin{figure}[ht]
  \centering
  \includegraphics[width=0.5\linewidth]{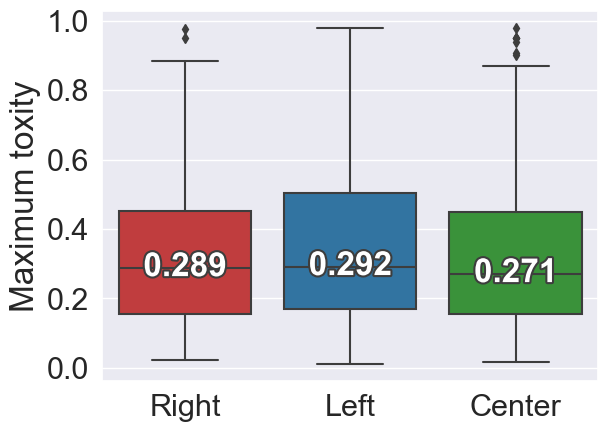}
  \caption{Boxplots represent the maximum toxicity of the replies for each user category. The median for each category is shown in each bar.
  }
  \label{fig2}
\end{figure}

In general, more frequently replied-to users shared lower maximum toxicities, regardless of user category.
The Left category differs from the others, possibly due to the higher toxicity values of several outliers.
For instance, some Left category outliers (indicated by arrows in Figure \ref{fig1}) shared larger toxicities and some of them even reached more than $0.8$.
The outliers could be top toxic repliers.
By contrast, the Right category users' maximum toxicities were less than $0.4$, when they were replied  more than, approximately 1,000 times.
The maximum toxicity is compared between categories in
Figure \ref{fig2}.
This reveals that the maximum toxicities of the Left category users are significantly higher than those of the other two categories (Mann-Whitney U test, $p < 0.005$).

\subsection{The Left and Center outliers received much more toxic replies.}

The median can be used to represent the centre tendency of a dataset.
In contrast to the maximum scenario, the level of median toxicity did not exhibit a negative correlation with replied times.

\begin{figure}[H]
  \centering
  \includegraphics[width=0.5\linewidth]{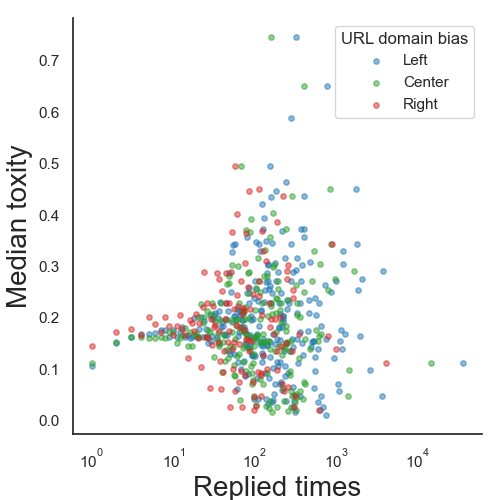}
  \caption{The median toxicities of replies of each politically-leaning category of replied-to users received. The three categories shared a similar distribution for median toxicity. The X-axis, repied times, is in log scale. The Y axis indicates the times each category of users were replied. Reds indicate users were engaging with Right domains, exclusively; Greens indicate users were engaging with Center domains, exclusively; Blues indicate uses were engaging with Left domains, exclusively. The Left and Center outlier users (indicated by arrows) received much more toxic replies. 
  }
  \label{fig3}
\end{figure}

\begin{figure}[H]
  \centering
  \includegraphics[width=0.5\linewidth]{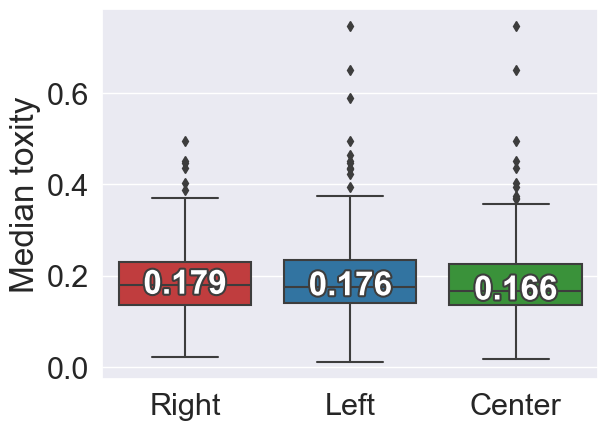}
  \caption{Boxplots represent the median toxicity of the replies for each user category. The median for each category is shown in each bar. The overall reply toxicities were similar across the three categories, although the Left and Center outliers received much more toxic contents.
  }
  \label{fig4}
\end{figure}

Most of the median toxicity values were concentrated between around $0.05$ to $0.4$.
This overall tendency showed that the toxicity of replies was less aggressive, but fluctuated as the replied times increased.
Specifically, when we looked at the Right category users, the median toxicities were below $0.5$, but the outlier values for Left and Center users reached over $0.7$.
No statistical significance was identified across the Left, Right, and Center, suggesting the three categories shared a similar distribution for median toxicity (Figure \ref{fig3}), and no significant median toxicity group was identified out of the three categories (Figure \ref{fig4}).

\section{Discussion}
This study shows that Left users could receive more toxic replies than Right and Center users.
This pattern of toxicity propagation is important for understanding misinformation propagation and echo chamber development, as toxicity in online interactions can lead to a decrease in user activity, ultimately impacting the collaborative nature of platforms ~\cite{10.1145/3366423.3380074}.
Previous research confirmed that the left group was more distant from the neutral group than the right group \cite{Miyazaki2022}. 
However, this study found that Left users were much closer to Right users than the Center user, in terms of maximum toxicities.
This ``toxicity distance'' might suggest that right and left users were sending toxicities to each other, but Left users received much more.  
Although there was no significant difference in the language toxicity across the replied-to users of the Left, Right, and Center categories, the replied users targeted by toxic repliers in each category cannot be neglectable, especially the Left users.

What precautions are necessary to take for protecting users from language toxicity attacks, especially during political discussions, such as U.S. presidential elections should be carefully considered.
When users are engaging the Left, it is suggested to pay attention to the toxic comments and replies, which might further pollute the SNS ecosystem and make users more emotional.
Future work would be finding out the dynamics of interaction and engagement dynamics for the Left, Right and Center. In addition, more intelligent tools need to be proposed to combat the aggression of the toxic language to keep our SNS ecosystem healthier.
This study has implications for other platforms, such as Facebook and Reddit.

\bibliographystyle{unsrt}  
\bibliography{references}  

\end{document}